\pdfoutput=1
\documentclass[conference]{IEEEtran}
\IEEEoverridecommandlockouts
\usepackage{cite}
\usepackage{amsmath,amssymb,amsfonts}
\usepackage{algorithmic}
\usepackage{graphicx}
\usepackage{textcomp}
\usepackage{xcolor}
\usepackage[hidelinks]{hyperref}
\usepackage{multicol}
\usepackage{multirow}

\usepackage{scalerel}
\usepackage{tikz}
\usetikzlibrary{svg.path}
\definecolor{orcidlogocol}{HTML}{A6CE39}
\tikzset{
	orcidlogo/.pic={
		\fill[orcidlogocol] svg{M256,128c0,70.7-57.3,128-128,128C57.3,256,0,198.7,0,128C0,57.3,57.3,0,128,0C198.7,0,256,57.3,256,128z};
		\fill[white] svg{M86.3,186.2H70.9V79.1h15.4v48.4V186.2z}
		svg{M108.9,79.1h41.6c39.6,0,57,28.3,57,53.6c0,27.5-21.5,53.6-56.8,53.6h-41.8V79.1z M124.3,172.4h24.5c34.9,0,42.9-26.5,42.9-39.7c0-21.5-13.7-39.7-43.7-39.7h-23.7V172.4z}
		svg{M88.7,56.8c0,5.5-4.5,10.1-10.1,10.1c-5.6,0-10.1-4.6-10.1-10.1c0-5.6,4.5-10.1,10.1-10.1C84.2,46.7,88.7,51.3,88.7,56.8z};
	}
}

\newcommand\orcidicon[1]{\href{https://orcid.org/#1}{\mbox{\scalerel*{
				\begin{tikzpicture}[yscale=-1,transform shape]
				\pic{orcidlogo};
				\end{tikzpicture}
			}{|}}}}

\usepackage{graphicx}
\usepackage{gensymb}
\usepackage{xcolor,soul}
\definecolor{myhlcolor}{gray}{0.85}
\sethlcolor{yellow}
\definecolor{myOrange}{rgb}{1,0.5,0}
\definecolor{myGreen}{rgb}{.2,.9,.2}

\def\BibTeX{{\rm B\kern-.05em{\sc i\kern-.025em b}\kern-.08em
    T\kern-.1667em\lower.7ex\hbox{E}\kern-.125emX}}
\begin{document}

\title{Cognitive Internet of Vehicles: Motivation, Layered Architecture and Security Issues\\
	{\footnotesize}
	\thanks{*Corresponding Author Email: k.fidahasan@yahoo.com}
}
\author{\IEEEauthorblockN{Khondokar Fida Hasan \orcidicon{0000-0002-8008-8203}*}
\IEEEauthorblockA{\textit{School of EECS, QUT} \\
Brisbane, Australia \\
}
\and
\IEEEauthorblockN{Tarandeep Kaur}
\IEEEauthorblockA{\textit{School of EECS, QUT} \\
Brisbane, Australia 
}
\and
\IEEEauthorblockN{Md. Mhedi Hasan}
\IEEEauthorblockA{\textit{Dept. of ICT, CoU} \\
Cumilla, Bangladesh 
}
\and
\IEEEauthorblockN{Yanming Feng}
\IEEEauthorblockA{\textit{School of EECS, QUT} \\
Brisbane, Australia 
}
}

\maketitle

\begin{abstract}
Over the past few years, we have experienced great technological advancements in the information and communication field, which has significantly contributed to reshaping the Intelligent Transportation System (ITS) concept.  Evolving from the platform of a collection of sensors aiming to collect data, the data exchanged paradigm among vehicles is shifted from the local network to the cloud. With the introduction of cloud and edge computing along with ubiquitous 5G mobile network, it is expected to see the role of Artificial Intelligence (AI) in data processing and smart decision imminent. So as to fully understand the future automobile scenario in this verge of industrial revolution 4.0, it is necessary first of all to get a clear understanding of the cutting-edge technologies that going to take place in the automotive ecosystem so that the cyber-physical impact on transportation system can be measured. CIoV, which is abbreviated from Cognitive Internet of Vehicle, is one of the recently proposed architectures of the technological evolution in transportation, and it has amassed great attention.  It introduces cloud-based artificial intelligence and machine learning into transportation system. What are the future expectations of CIoV? To fully contemplate this architecture’s future potentials, and milestones set to achieve, it is crucial to understand all the technologies that leaned into it. Also, the security issues to meet the security requirements of its practical implementation. Aiming to that, this paper presents the evolution of CIoV along with the layer abstractions to outline the distinctive functional parts of the proposed architecture. It also gives an investigation of the prime security and privacy issues associated with technological evolution to take measures.
\end{abstract}


\section{Introduction}
The transport sector plays a key role in modern civilisation, and over the past years, it has experienced rapid growth. According to a recent survey; 2019’s Motor Vehicle Census, there is an annual increase rate of 1.7\% (average) in Australia, with 19.5 million registered motor vehicles over a population of 25 million people \cite{vehiclecensus2019}. This number for the USA and the UK stands at 281.3 million \cite{usstatistices} and 39.4 million \cite{censusUK} respectively. In turns, the whole world has over 1.4 billion registered vehicles and come the year 2040, this figure is expected to shoot by double \cite{worldecoforum}. The critical issue is, as the number of vehicles on roads increases, traffic-related problems such as traffic congestion, accident, and road fatalities are on the rise as well. To counter-attack this, a smart transport and traffic management system was envisioned since the year 1990 by combining different sensors, and different mode of applications under the technological evolution of Intelligent Transportation System (ITS). At its earlier stage, this technological endeavour evolved with the integration of co-employable and assisting correspondence innovation termed as Co-operative Intelligent Transportation System (CITS), which basically incorporates the Information and Communication Technologies (ICT) with transportation infrastructure \cite{Sladkowski2016}. The vehicle is enabled to “communicate” with other vehicles, roadside infrastructures and other road entities by creating vehicular ad-hoc networks (VANET). However, both of these two technological advancements lead the concept of Autonomous and Connected vehicular technologies in parallel to enrich the idea of intelligence in the transport sector so as to increase the user comfort and road safety. 

With the evolution of the Internet of Things (IoT), in the early 2010s, meanwhile, vehicles are being connected to the Internet, aiming at providing ubiquitous access to information alike to the drivers and passengers. This leads to another technological break-through termed as Internet of Vehicle (IoV). 

\begin{figure}[t]
	\centering
	\includegraphics[height=0.25\textheight, width=0.5\textwidth]{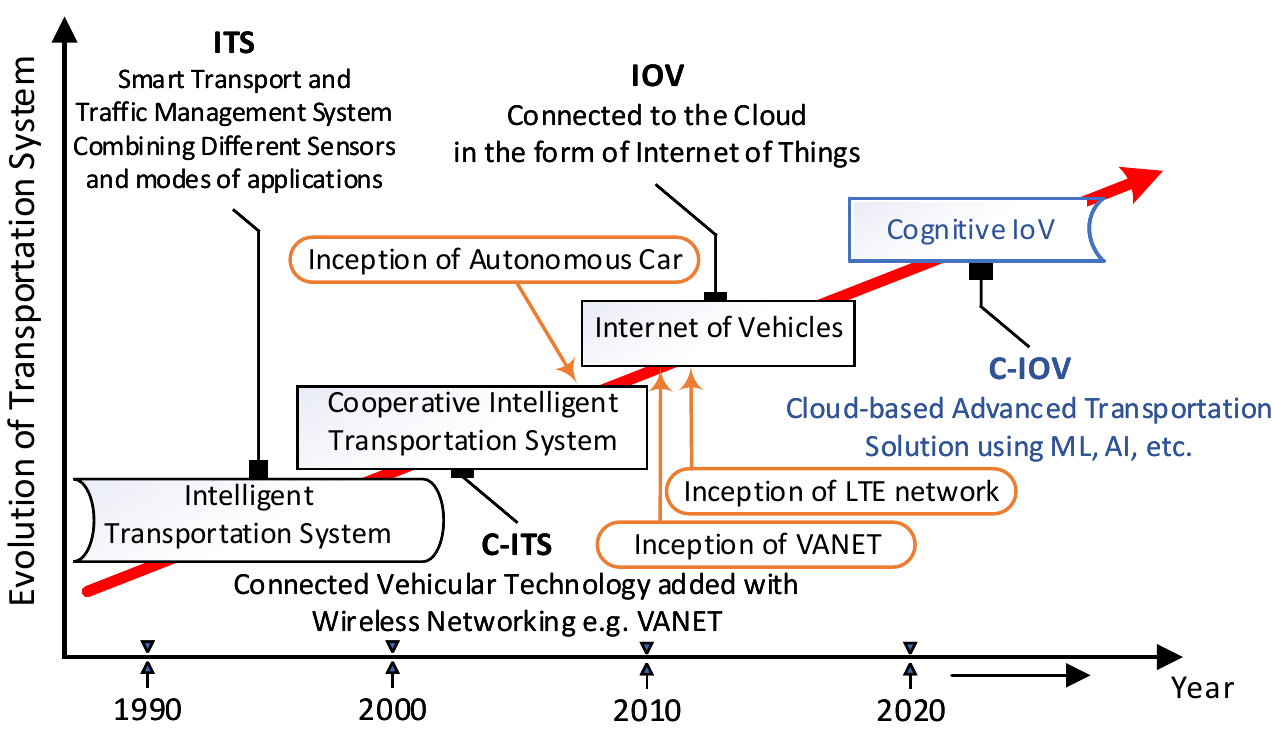}
	\caption{\small Evolution of Cognitive Internet of Vehicles (CIoV).}
	\label{Figure1}
\end{figure}

Even though there has been a noticeable advancement in terms of automation and connectivity, still, it is not sufficient to reduce the road causalities to zero. Driving errors, as well as the drivers’ misjudgement being the prime reasons associated with this causalities where a recent research \cite{Singh2015} reveals that 90\% of road accidents are caused by human factors. Take, for instance, fatigue while driving, overspreading, blocked line of sight, etc., are ranked among the most common factors that cause accidents. This encourages the necessity of applying Machine Learning (ML), Neural Network (NN), Deep Learning (DL), and Artificial Intelligence (AI) that can take control of wheel which can enable error-free driving, resulting to the idea of Cognitive Internet of Vehicles (CIoV) shown in \figurename{\ref{Figure1}}.

The advancement that the CIoV offers toward the use of internet and machine intelligence are also associated with new security risks and privacy issues, realising the transportation need to address properly. Since different technologies are playing a role in different layers, it is vital to understand the existing vulnerabilities in the generic domain of those technologies and their application. 

This paper aims at giving an overview of the evolution of Cognitive Internet of Vehicles (CIoV) and its technological related reviews. It presents a five-layer model to envisage the architecture of future transportation system to identify their distinctive functional parts. This paper also discusses security risks, including different threats, attacks, and vulnerabilities may associate with different layers to understand the measures required. 

\section{Cognitive Internet of Vehicles (CIoV) and Layered Architecture}
With the advent of the ever-growing vehicular applications, the technical challenges are growing too to meet the demands from both communication and computation. Without persuasive communication and computational support, a good number of foreseeing vehicular applications and services will only still stay in the idea phase and cannot be seen into practice. 

\subsection{Sensing and Participation}
The Layer-1 of the structure represents all the technologically evolved entities that are capable of sensing and communicating and also responsible for interacting within the transportation system. Such entity includes smart vehicles and road infrastructure. Smart vehicles are generally considered to have a certain level of intelligence. With the variety of technologies, smart vehicles can be categorised into two, autonomous vehicles and connected vehicles. Autonomous vehicles are generally referred to as the driverless vehicle. The involvement of the driver while driving and operating Autonomous vehicles depends on the levels of automation, ranges from level 0 to level 4. Level-0 indicates no automation where the driver controls the vehicle at all the times including the steering, braking, etc. solely handled by a human being. Whereas, Level-4 refers to a fully self-driving, self-operated automobile. In such a level of automation, the vehicle operates on its own without any human assistance \cite{anderson2014autonomous,Lu2019}. In between these two levels, the other three levels show some degree of automation in their operations but not entirely at all. 
Autonomous Vehicles are aligned to work on a three-phase design known as sense-plan-act, which is the base of many robotic systems. Mostly, all the Automated vehicles are equipped with a number of sensors, cameras, Lidar, Radar, etc., that collects raw data from the external environment. This data then serves as the input to the sophisticated system software which is used in vehicles to decide for specific courses of actions, such as, lane changing, acceleration and overtaking other vehicles \cite{bagloee2016autonomous}. In connected vehicular technology, however, the vehicles communicate with internal and external environments utilising a different kind of communication technologies predominantly wireless communication technologies. These vehicles use wireless networks to create interactions within the devices built in the vehicle itself that is On-Board sensors and outside the vehicle; that is Vehicle to Vehicle (V2V) communications or Vehicle to Infrastructure (V2I) communications \cite{lu2014connected}. The concept of connected technologies fundamentally propelled the evolution of the transportation system to form the Internet of Vehicles (IoV), thus creating an opportunity to apply modern technological developments to apply on data such as machine learning and artificial intelligence, to creates insights on transportation management and to take measures on providing better services.

\subsection{Network Communication and Data Acquisition}
\begin{figure*}[!h]
	\begin{center}
		\includegraphics[height=0.30\textheight, width=0.85\textwidth]{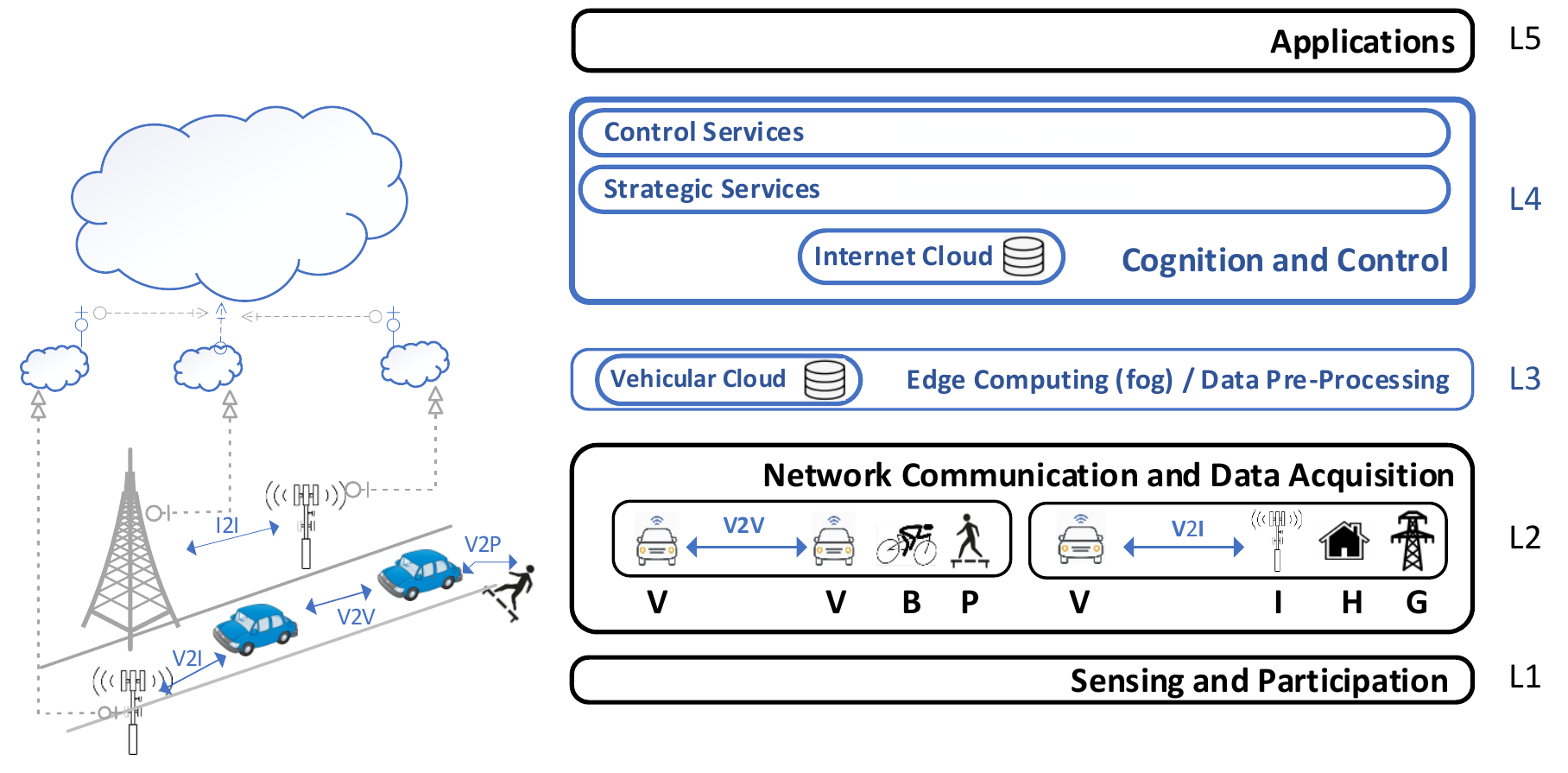}
	\end{center}
	\caption{\small Layered Architecture of Cognitive Internet of Vehicles (CIoV).}
	\label{Figure2}
\end{figure*}

Layer-2 in CIoV is responsible for network-based communications among different transportation entities aiming at the transport-related data acquisition. A wide variety of communication interaction that takes place are defined in this layer. Broadly, all interaction can be classified as Intra-vehicular communication and Inter-vehicular communications. Communication that takes place within the vehicle is termed as Intra-vehicular communication. Generally, smart vehicles are equipped with numerous sensors, such as sensors detect the road condition, drivers’ fatigue, monitoring of the tire pressure, and autonomous control sensors, etc. \cite{lu2014connected}. The primary objectives of those sensors in vehicles are to monitor the internal operation of vehicle. Those sensors communicate with each other and take intelligent decision for the human driver. Smart vehicles use technologies that allow them to make decisions for the driver. For example, crash warning systems, lane changing systems, adaptive cruise control, and self-parking are some of the examples that operate using the concept of intra-vehicular sensor-based communications \cite{anderson2014autonomous}. 
Inter-vehicular communication, however, can be categorised as Vehicle to Vehicle (V2V) including Device (V2D), Pedestrian (V2P), Bicycle (V2B), etc. and Vehicle to Infrastructure (V2I) including Home (V2H), Smar Grid (V2G), etc. as shown in the \figurename{\ref{Figure2}}.


\subsection{Edge Computing and Data Pre-Processing}
This layer forms on top of the vehicular cloud to support all sorts of computing services at the edge of the network. It aims at collecting data from lower layers to support storing and processing. The layer is also responsible for providing real-time services to the participating agents \cite{hou2016vehicular}. Such vehicular cloud concept helps provide users with answers to location-based real-time queries faster. For example, a driver dispatches the query regarding sudden traffic jam, so the answer to this query is provided by vehicular cloud. Some of the other real-time services that can be offered by the vehicular cloud include; Navigation, Crash warnings system, Traffic Monitoring, Parking Availability, Autonomous Driving, etc. The vehicular cloud serves is used with such kind of real-time services. The main driving force behind the formation of vehicular clouds is that in the near future, a huge number of vehicles is expected on the roads, streets and parking lots and these vehicles can be assisted with computational and communication resources from the edge units support from the form of fog computing, that will help to manage the traffic better and reduce the cost that are incurred in managing roads and traffic with traditional approaches \cite{Hussain2012}.


	\subsection{Cognition and Control}
\begin{figure}[!h]
	\centering
	\includegraphics[height=0.28\textheight, width=0.5\textwidth]{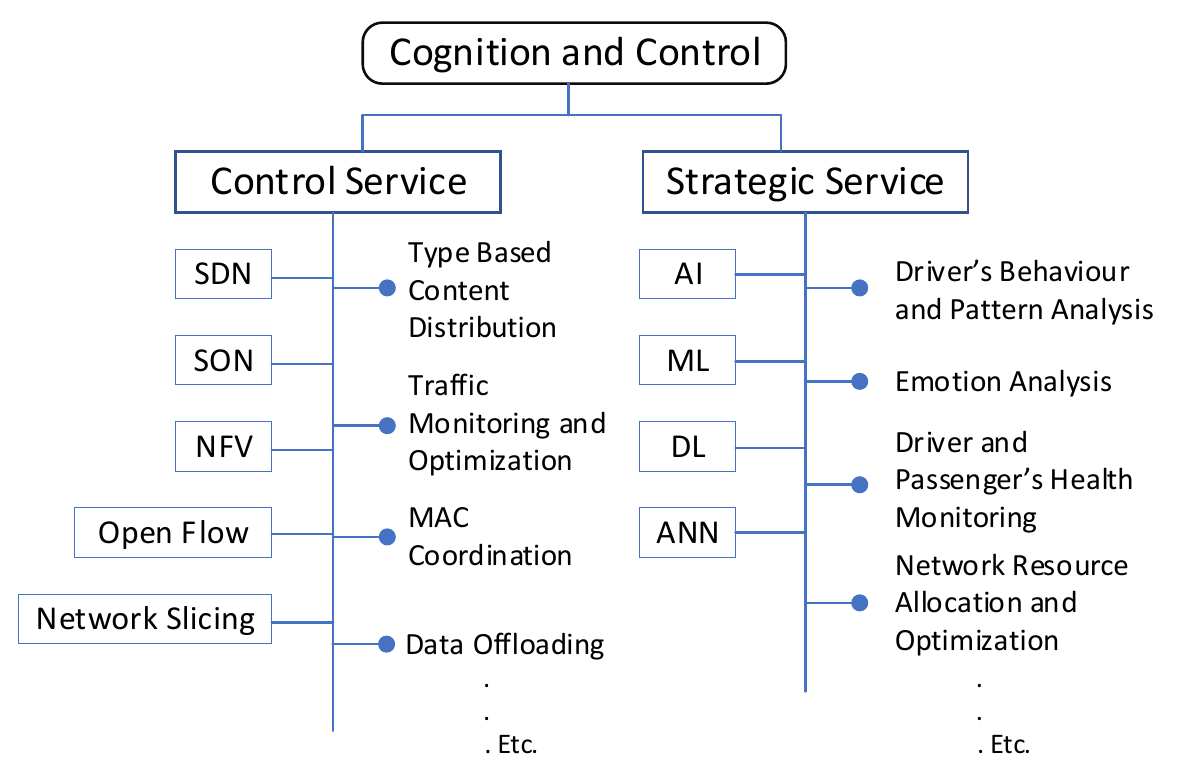}
	\caption{\small Enabling Engines and Functions of the Layer Cognition and Control in CIoV Realisation \cite{kaiwartya2016internet,chen2018cognitive,Bhatia2019}.}
	\label{Figure3}
\end{figure}

Cognition and Control layer forms in the top of the internet cloud that fundamentally provides the feature of storing, computing and processing to the data collected from lower layers along with decision making support over it. Therefore, two different sorts of services can be distinctly considered in this layer; (i) Control service and (ii) Strategic service. In the strategic services sub-layer a variety of operations and analysis takes place, for example, driver's emotion and behaviour analysis, driver and passenger's health condition monitoring, different road condition monitoring, and network optimization, etc. This is achieved through cloud-based dynamic cognition and utilisation of the computing resources by enabling advanced operations using machine learning, deep learning, neural networking, and artificial intelligence which essentially sort of data cognitive engine process to perform better analysis of the heterogeneous data. The control sub-part of Cognition and Control layer, however, responsible to determine system performance by providing a verity of services such as network-wide security, traffic and network optimisation, and resource allocation through different resources management engine concept, for example, Software Defined Network (SDN), and Network Function Virtualisation (NFV), etc. \cite{kaiwartya2016internet,chen2018cognitive,Bhatia2019}. A list of enabling technologies and functions according to these two services is presented in \figurename{\ref{Figure3}}.

It is also worth to be mentioned that in this architecture this layer differs from the vehicular cloud as we think the internet cloud is public and is meant to serve all users of CIoV whereas the vehicular cloud is relevantly limited in some areas and serve the purpose of data computing, and storing in that restricted area.

\subsection{Application Layer}
The application layer provides high-level services that are a type of asynchronous functions can be treated as the end products of the architecture, help CIoV in fulfilling the objective of Intelligent transportation system in terms of driver assistance and congestion-free better traffic management. This layer also involves coordination and collaboration with different parties involved in transportation system such as automatic and mobile connected services.
Overall, with this paradigm shift, the Intelligent transportation system is leapt forward \cite{shankarwar2015security}. 

\section{Security and Privacy Based Layered Architecture}
\begin{figure*}[!h]
	\begin{center}
		\includegraphics[height=0.70\textheight, width=1\textwidth]{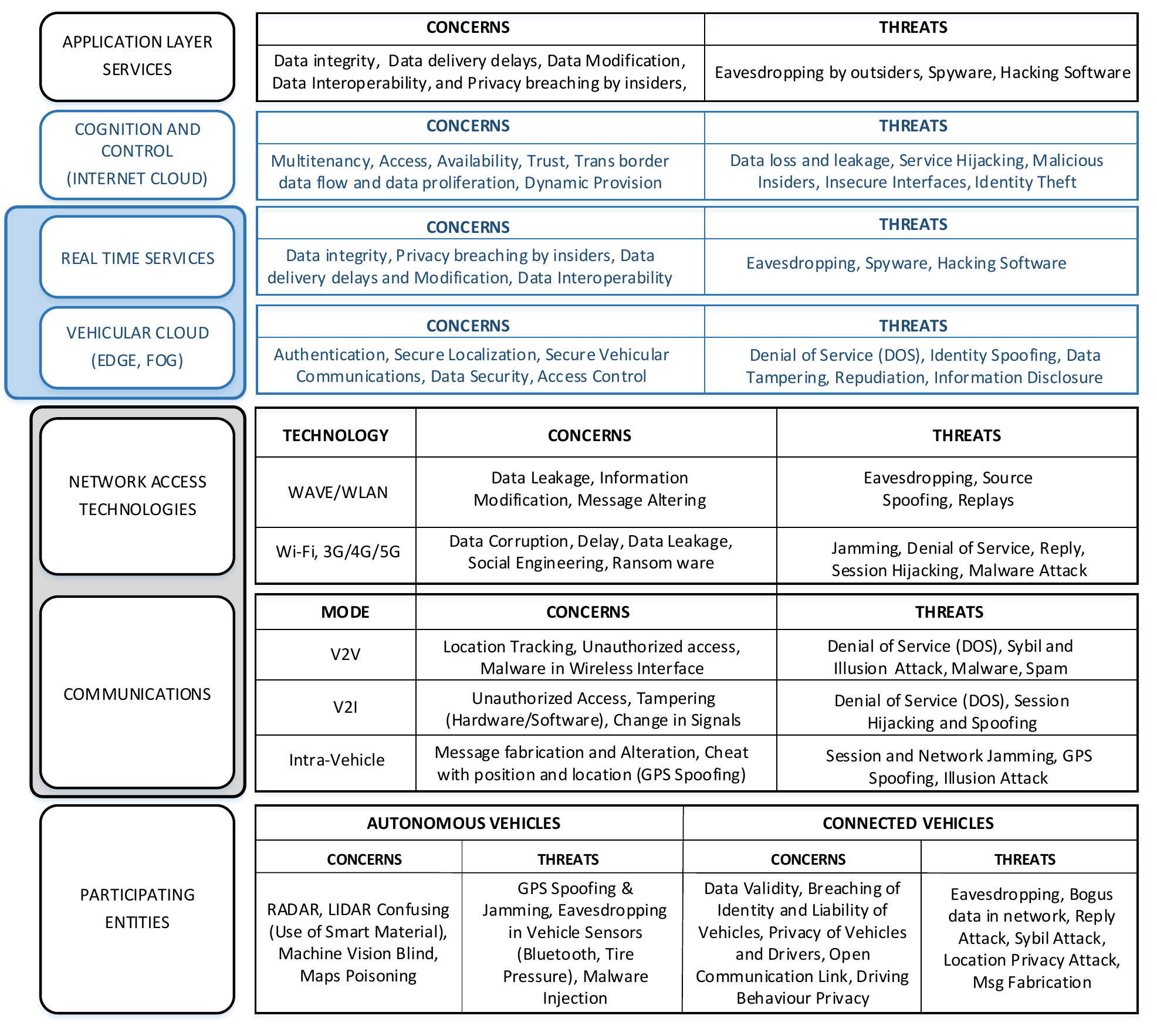}
	\end{center}
	\caption{\small Security Concerns and Threats model in CIoV layered Architecture.}
	\label{Figure4}
\end{figure*}

In this section, security and privacy are studied concerning the evolved technologies in the functioning of CIoV. Security is all about the safeguarding of data, and in transportation, security is a serious issue as it affects the lives of commuters using the roads directly.  For example, due to the lack of security measures, network intrusion can take place in vehicular network from outside the internet that may result in hijacking the vehicles by hackers. Privacy, however, is about safeguarding the user identity which is also equally important to be maximised by taking the relevant security measures.
Based on the architecture proposed above in \figurename{\ref{Figure2}}, security concerns and threats are investigated into a detailed seven-layered structure in \figurename{\ref{Figure4}}. Generally, the security and privacy view are presented as security concerns and threats. Threats are circumstances and events that may harm an information system through unauthorised access, destruction, disclosure, modification of data, and denial of service. Threats are performed by attackers who compromise systems resources \cite{li2018security}. The following sections focusses on the security, concerns and threats. 

\subsection{Sensing and Participating Entities}
As the layer has been defined, the Security and Privacy Issues of the Layer-1 are related to the security and privacy contexts of autonomous and connected vehicles. Although some of them are common, many of them are technology-specific, thus presented are separately in the layered diagram. First of all, in autonomous car technologies, the identified security issue is confusing the RADAR and LIDAR sensors which result in the vision of the machine going “blind”. This is caused by the attacks on cameras and sensors in vehicle and map poisoning, that is, altering the positions provided by a map of vehicle. The identified threats in autonomous technologies are eavesdropping in the in-vehicle sensors, entering of malware in in-vehicle system and GPS spoofing. Among them, GPS spoofing is a common threat that confuses the GPS of the vehicle by sending false strong signals by using other GPS simulators \cite{jadaan2017connected}.
In the case of connected vehicular technology, however, the security issues are related to the data validity, breaching of privacy of drivers, and open communication link \cite{othmane2015survey}. The data in connected vehicles includes; in-vehicle data, location data, and aggregated data. Using the communication links in connected vehicles, communication between On-board Unit and Roadside unit), attackers breach the privacy of drivers by getting access to the drivers’ private data which include location information, etc. There are several threats identified as well in the context of connected vehicles. According to \cite{othmane2015survey,jadaan2017connected}, the specific threats in connected vehicles are transferring bogus information to other drivers, changing the positioning information to ignore liability, performing a denial of service attacks to break down the network and tracking other vehicles by identifying them in-network. Reference \cite{jadaan2017connected} has also provided the security and privacy threats that occur in context of connected vehicles, these include, eavesdropping the communication between two parties, fabricating the messages, that is, creating the false signals and performing the Replay attacks, that is, modifying the timestamp of message and broadcasting it multiple times to disturb the traffic.

\subsection{Network Communication and Data Acquisition}
To understand better the security model of CIoV, this layer is presented here into two sub-layers utilising different communication paradigm and network technologies. While communication is a concern, the security and privacy issues along with threats, are identified by exploring the intra-vehicle communications and inter-vehicle communications, that is, V2V and V2I. The common security and privacy issues in all the communications are Tracking of vehicles locations, unauthorised access, tampering of hardware, Message alteration and fabrication \cite{7428337}. The identified threats are Denial of service (making network unavailable), Sybil attack, Session Hijacking (entering the communication session by cheating with the IP address of the device in vehicle), illusion attack (making vehicles to see false view) and GPS spoofing \cite{7428337,haus2017security,Hasan2018,shankarwar2015security,Hasan2018a}. However, under network access domain, the security and privacy issues identified include Data Leakage, Information Modification Message Altering, Data corruption. Threats are also a major concern that exists in the WAVE technology and these threats include: Eavesdropping, Source Spoofing, Wormhole attack, Denial of service, Replay and Jamming (attacker creates strong signals to disrupt the communications), are the threats that also occur in the Wi-Fi technology and cellular networks \cite{haus2017security}.

\subsection{Edge Computing and Data Pre-Processing}
This layer is responsible for providing real-time services to the participating entities based on the intelligent computation of data that is collected from communications of participating vehicles. Thus, for proper outlining, the investigated security concerns and threats are presented into two different sub-layers. Firstly, the security issues that occur in vehicular cloud includes Breaching of confidentiality, which means cheating with other users identities, valuable data and documents that are stored in vehicular cloud acquiring locations of vehicles from vehicular cloud Compromise with integrity, which include misuse and modification of valuable data of users that is stored in Vehicular cloud \cite{yan2012towards}. The security and privacy issues in real-time services, however, are identified by the reference \cite{wang2011real} in terms of privacy issues, authorisation and authentication of users. Privacy issues include breaching of privacy of users by acquiring their valuable data. Data integrity, Privacy breaching by insiders, Data delivery delays, Data Modification are the security issues in real-time services, also threats include eavesdropping, hacking software that provides the services and change the information that will be provided to end-users and spyware, in which attackers install software on other vehicles On-board units to get the positions of vehicle \cite{hoh2006enhancing}.

\subsection{Cognition and Control}
In this layer; Cognition and Control layer, there are a number of security and privacy issues identified, which are mostly related to the security issues and are a threat, to the Internet Cloud. According to Reference \cite{shankarwar2015security,li2018security,takabi2010security,joy2017internet}, some of the security issues that exist in the cloud paradigm include:
\subsubsection{Multitenancy} This is the capability of a cloud to run and operate on multiple machines, which eventually makes it more vulnerable to attacks on cloud infrastructure. 
\subsubsection{Access} So as to have access to sensitive user’s data, attackers may hack the system, thereby having access to the data store.
\subsubsection{Availability} The cloud is designed in such a way that users can access the available data at any time, irrespective of their geographic location. But what happens if there is a system failure? This could result in data failure, thus loss in the systems confidence by some users. 
\subsubsection{Misuse of Cloud computing} Users have unlimited access to network and storage while using cloud. In some cases, the cloud provider may give free trials, which could result in its misuse, thus adversely affecting cloud computing. 
\subsubsection{Transborder data flow and data proliferation} The data stored, can be accessed by selected companies, with or without the consent of the data owner. In such a case, it would be hard to ensure that the data in question is not stored or processed in some unauthorised systems.
\subsubsection{Trust} Most of the cloud users lack the total trust in it while storing their highly valued and private data, with the fear that the systems may collapse, thus losing their data. The threats are also identified in the cloud computing layer. 
The threats, however, includes Service hijacking, Identity theft, Malicious insiders and Data loss, and leakage are some common threats existing in the cloud domain \cite{modi2013survey}.

\subsection{Application Layer}
The security and privacy issues are dependent on all the lower layers, the main security concern in this layer can be the delay in reaching of data to the applications or software that are responsible for providing high-level services such as reduced rate of accidents and congestions. The second security concern is Data interoperability, which means that data coming from different sources from lower layers may be in an incompatible format for the application and software that provide end services \cite{othmane2015survey}.

\section{Conclusion}
In this paper, we present a five-layer based novel transportation architecture for the future automobile industry. With the growing technological trend, the proposed structure embeds internet-based cognitive intelligence and also explains the functions of all the layer abstractions. This cover outlines this novel’s architecture of CIoV, in order to indicate research opportunities in a vehicular network. This paper also investigates and focuses on the security and privacy issues identified along with the proposed architecture layer by layer. The concept along with the investigated result on security and privacy will help transition to the Cloud-based future transportation, providing all the security services may be required by the autonomous vehicles.

\bibliographystyle{IEEEtran}
\bibliography{myref}

\begin{thebibliography}{10}
\providecommand{\url}[1]{#1}
\csname url@samestyle\endcsname
\providecommand{\newblock}{\relax}
\providecommand{\bibinfo}[2]{#2}
\providecommand{\BIBentrySTDinterwordspacing}{\spaceskip=0pt\relax}
\providecommand{\BIBentryALTinterwordstretchfactor}{4}
\providecommand{\BIBentryALTinterwordspacing}{\spaceskip=\fontdimen2\font plus
\BIBentryALTinterwordstretchfactor\fontdimen3\font minus
  \fontdimen4\font\relax}
\providecommand{\BIBforeignlanguage}[2]{{%
\expandafter\ifx\csname l@#1\endcsname\relax
\typeout{** WARNING: IEEEtran.bst: No hyphenation pattern has been}%
\typeout{** loaded for the language `#1'. Using the pattern for}%
\typeout{** the default language instead.}%
\else
\language=\csname l@#1\endcsname
\fi
#2}}
\providecommand{\BIBdecl}{\relax}
\BIBdecl

\bibitem{vehiclecensus2019}
\BIBentryALTinterwordspacing
A.~B. of~Statistics. (2019, Jan.) Motor vehicle census, australia. Australian
  Bureau of Statistics. [Online]. Available:
  \url{https://www.abs.gov.au/ausstats/abs@.nsf/mf/9309.0}
\BIBentrySTDinterwordspacing

\bibitem{usstatistices}
\BIBentryALTinterwordspacing
U.~{S}tatistics. {US} {VIO} {V}ehicle {R}egistration {D}ata 2018, {F}ast
  {Q}uote on {C}ar {D}ata. [Online]. Available:
  \url{https://hedgescompany.com/automotive-market-research-statistics/auto-mailing-lists-and-marketing/}
\BIBentrySTDinterwordspacing

\bibitem{censusUK}
\BIBentryALTinterwordspacing
A.~{P}ublishing {S}ervice. (2019, Apr.) Vehicle {L}icensing {S}tatistics:
  Annual 2018. [Online]. Available:
  \url{https://assets.publishing.service.gov.uk/government/uploads/system/uploads/attachment_data/file/800502/vehicle-licensing-statistics-2018.pdf}
\BIBentrySTDinterwordspacing

\bibitem{worldecoforum}
\BIBentryALTinterwordspacing
W.~{E}conomic {F}orum. (2016, Sep.) The number of cars worldwide is set to
  double by 2040. Accessed on. [Online]. Available:
  \url{https://www.weforum.org/agenda/2016/04/the-number-of-cars-worldwide-is-set-to-double-by-2040}
\BIBentrySTDinterwordspacing

\bibitem{Sladkowski2016}
A.~S{\l}adkowski and W.~Pamu{\l}a, \emph{Intelligent transportation
  systems-problems and perspectives}.\hskip 1em plus 0.5em minus 0.4em\relax
  Springer, 2016, vol. 303.

\bibitem{Singh2015}
S.~Singh, ``Critical reasons for crashes investigated in the national motor
  vehicle crash causation survey,'' Tech. Rep., 2015.

\bibitem{anderson2014autonomous}
J.~M. Anderson, K.~Nidhi, K.~D. Stanley, P.~Sorensen, C.~Samaras, and O.~A.
  Oluwatola, \emph{Autonomous vehicle technology: A guide for
  policymakers}.\hskip 1em plus 0.5em minus 0.4em\relax Rand Corporation, 2014.

\bibitem{Lu2019}
H.~Lu, Q.~Liu, D.~Tian, Y.~Li, H.~Kim, and S.~Serikawa, ``The {C}ognitive
  {I}nternet of {V}ehicles for {A}utonomous {D}riving,'' \emph{IEEE Netw.},
  vol.~33, no.~3, pp. 65--73, 2019.

\bibitem{bagloee2016autonomous}
S.~A. Bagloee, M.~Tavana, M.~Asadi, and T.~Oliver, ``Autonomous {V}ehicles:
  {C}hallenges, {O}pportunities, and {F}uture {I}mplications for
  {T}ransportation {P}olicies,'' \emph{Jo. Mod. Trans.}, vol.~24, no.~4, pp.
  284--303, 2016.

\bibitem{lu2014connected}
N.~Lu, N.~Cheng, N.~Zhang, X.~Shen, and J.~W. Mark, ``Connected vehicles:
  {S}olutions and {C}hallenges,'' \emph{IEEE Internet Things J.}, vol.~1,
  no.~4, pp. 289--299, 2014.

\bibitem{hou2016vehicular}
X.~Hou, Y.~Li, M.~Chen, D.~Wu, D.~Jin, and S.~Chen, ``Vehicular fog computing:
  {A} viewpoint of vehicles as the infrastructures,'' \emph{IEEE T VEH
  TECHNOL}, vol.~65, no.~6, pp. 3860--3873, 2016.

\bibitem{Hussain2012}
R.~Hussain, J.~Son, H.~Eun, S.~Kim, and H.~Oh, ``Rethinking {V}ehicular
  {C}ommunications: {M}erging {VANET} with {C}loud {C}omputing,'' in \emph{4th
  IEEE Inter. Confer. on Cloud Computing Tech. Sci}.\hskip 1em plus 0.5em minus
  0.4em\relax IEEE, 2012, pp. 606--609.

\bibitem{kaiwartya2016internet}
O.~Kaiwartya, A.~H. Abdullah, Y.~Cao, A.~Altameem, M.~Prasad, C.-T. Lin, and
  X.~Liu, ``Internet of vehicles: Motivation, layered architecture, network
  model, challenges, and future aspects,'' \emph{IEEE Access}, vol.~4, pp.
  5356--5373, 2016.

\bibitem{chen2018cognitive}
M.~Chen, Y.~Tian, G.~Fortino, J.~Zhang, and I.~Humar, ``Cognitive {I}nternet of
  {V}ehicles,'' \emph{Com Com. J.}, vol. 120, pp. 58--70, 2018.

\bibitem{Bhatia2019}
J.~Bhatia, Y.~Modi, S.~Tanwar, and M.~Bhavsar, ``Software defined vehicular
  networks: A comprehensive review,'' \emph{International Journal of
  Communication Systems}, vol.~32, no.~12, p. e4005, 2019.

\bibitem{shankarwar2015security}
M.~U. Shankarwar and A.~V. Pawar, ``Security and {P}rivacy in {C}loud
  {C}omputing: {A} {S}urvey,'' in \emph{Pro. 3rd Interl Conf. on Front. of
  Intelli. Com. Theory and App. (FICTA) 2014}.\hskip 1em plus 0.5em minus
  0.4em\relax Springer, 2015, pp. 1--11.

\bibitem{li2018security}
H.~Li, R.~Lu, J.~Misic, and M.~Mahmoud, ``Security and {P}rivacy of {C}onnected
  {V}ehicular {C}loud {C}omputing,'' \emph{IEEE Net.}, vol.~32, no.~3, pp.
  4--6, 2018.

\bibitem{jadaan2017connected}
K.~Jadaan, S.~Zeater, and Y.~Abukhalil, ``Connected {V}ehicles: {A}n
  {I}nnovative {T}ransport {T}echnology,'' \emph{Procedia Engineering}, vol.
  187, pp. 641--648, 2017.

\bibitem{othmane2015survey}
L.~B. Othmane, H.~Weffers, M.~M. Mohamad, and M.~Wolf, ``A survey of security
  and privacy in connected vehicles,'' in \emph{Wireless sensor and mobile
  ad-hoc networks}.\hskip 1em plus 0.5em minus 0.4em\relax Springer, 2015, pp.
  217--247.

\bibitem{7428337}
Y.~{Sun}, L.~{Wu}, S.~{Wu}, S.~{Li}, T.~{Zhang}, L.~{Zhang}, J.~{Xu}, and
  Y.~{Xiong}, ``Security and {P}rivacy in the {I}nternet of {V}ehicles,'' in
  \emph{2015 Inter. Conf. on Identi., Infor, and Knowl in the Internet of
  Things (IIKI)}, Oct 2015, pp. 116--121.

\bibitem{haus2017security}
M.~Haus, M.~Waqas, A.~Y. Ding, Y.~Li, S.~Tarkoma, and J.~Ott, ``Security and
  {P}rivacy in {D}evice-to-{D}evice (d2d) {C}ommunication: {A} {R}eview,''
  \emph{IEEE Commun. Surv. Tutor.}, vol.~19, no.~2, pp. 1054--1079, 2017.

\bibitem{Hasan2018}
K.~F. Hasan, Y.~Feng, and Y.-C. Tian, ``{GNSS} {T}ime {S}ynchronization in
  {V}ehicular {A}d-hoc {N}etworks: {B}enefits and {F}easibility,'' \emph{IEEE
  Trans. Intell. Transp. Syst.}, vol.~19, no.~12, pp. 3915--3924, 2018.

\bibitem{Hasan2018a}
K.~F. Hasan, C.~Wang, Y.~Feng, and Y.-C. Tian, ``Time {S}ynchronization in
  {V}ehicular {A}d-hoc {N}etworks: {A} {S}urvey on {T}heory and {P}ractice,''
  \emph{Veh. Commun.}, vol.~14, pp. 39--51, 2018.

\bibitem{yan2012towards}
G.~Yan, D.~B. Rawat, and B.~B. Bista, ``Towards {S}ecure {V}ehicular
  {C}louds,'' in \emph{2012 Sixth Inter. Conf. on Complex, Intelligent, and
  Software Intensive Systems}.\hskip 1em plus 0.5em minus 0.4em\relax IEEE,
  2012, pp. 370--375.

\bibitem{wang2011real}
J.~Wang, J.~Cho, S.~Lee, and T.~Ma, ``Real time services for future cloud
  computing enabled vehicle networks,'' in \emph{2011 Int. conf. on wireless
  com. and signal processing (WCSP)}.\hskip 1em plus 0.5em minus 0.4em\relax
  IEEE, 2011, pp. 1--5.

\bibitem{hoh2006enhancing}
B.~Hoh, M.~Gruteser, H.~Xiong, and A.~Alrabady, ``Enhancing security and
  privacy in traffic-monitoring systems,'' \emph{IEEE Pervasive Comput.},
  vol.~5, no.~4, pp. 38--46, 2006.

\bibitem{takabi2010security}
H.~Takabi, J.~B. Joshi, and G.-J. Ahn, ``Security and privacy challenges in
  cloud computing environments,'' \emph{IEEE Secur. Priv.}, vol.~8, no.~6, pp.
  24--31, 2010.

\bibitem{joy2017internet}
J.~Joy and M.~Gerla, ``Internet of vehicles and autonomous connected
  car-privacy and security issues,'' in \emph{2017 26th Inter. Conf. on Comp.
  Commun. and Net. (ICCCN)}.\hskip 1em plus 0.5em minus 0.4em\relax IEEE, 2017,
  pp. 1--9.

\bibitem{modi2013survey}
C.~Modi, D.~Patel, B.~Borisaniya, A.~Patel, and M.~Rajarajan, ``A survey on
  security issues and solutions at different layers of cloud computing,''
  \emph{The j supercomputing}, vol.~63, no.~2, pp. 561--592, 2013.

\end{thebibliography}



\end{document}